\documentstyle[prb,aps,amssymb,psfig,floats]{revtex}
\begin{document}
%\draft
%\preprint{HEP/123-qed}
%\tighten
%
\title{Exactly solvable two-dimensional quantum spin models}
\author{D. V. Dmitriev, V. Ya. Krivnov, and A. A. Ovchinnikov}
\address{Joint Institute of Chemical Physics, Russian
Academy of Sciences, 117977~Moscow, Russia}
\date{JETP {\bf 88}, 138 (1999)}
%Zh. \'{E}ksp. Teor. Fiz. {\bf 115}, 249--267 (January 1999}
\maketitle
\begin{abstract}
A method is proposed for constructing an exact ground-state wave function
of a two-dimen\-sional model with spin~1/2.  The basis of the method is to
represent the wave function by a product of fourth-rank spinors associated
with the sites of a lattice and the metric spinors corresponding to bonds
between nearest neighbor sites.  The function so constructed is an exact
wave function of a 14-parameter model.  The special case of this model
depending on one parameter is analyzed in detail.  The ground state is
always a nondegenerate singlet, and the spin correlation functions decay
exponentially with distance.  The method can be generalized for models with
spin~1/2 to other types of lattices.
\end{abstract}

%\narrowtext

\renewcommand{\thesection}{\arabic{section}}

\section[]{Introduction}\label{s1}

There has been growing interest lately in quantum spin systems with
frustrated interactions.\cite{1,2,3,4,5,6,7,8,9,10}  Of special importance are
models in this category for which it is possible to construct an exact
ground state.  The first example of such a representation is the well-known
Majumdar-Ghosh model.\cite{11}  It comprises a chain of spins~1/2 with
antiferromagnetic interactions $J_{1}$ and $J_{2}$ of nearest neighbor and
next-nearest neighbor spins, where $J_{2} = J_{1}/2$.  The ground state of
this model is two-fold degenerate and consists of dimerized singlets;
moreover, there is a gap in the spectrum of excitations.  Another example
of an exactly solvable model is the one-dimen\-sional model with bilinear
and biquadratic interactions and spin~1, investigated by Affleck, Kennedy,
Lieb, and Tasaki\cite{12} (AKLT model).  Its ground state has a structure of
the type where each neighboring pair of spins has valence bonds.  It is not
degenerate, the spin correlation functions in the ground state decrease
exponentially with distance, and there is a gap in the spectrum of
excitations.  This model therefore has properties predicted by Haldane\cite{13}
for the one-dimen\-sional Heisenberg antiferromagnetic model with spin~1.
The valence-bond ground state is also exact for systems with many
dimensions, but with spin $d/2$ ({\it d} is the coordination number of the
lattice).\cite{14}  The one-dimen\-sional AKLT model has subsequently been
generalized and investigated in a number of papers,\cite{15} where it has also
been shown that the wave function of the ground state can be represented by
the trace of the product of matrices describing the spin states of sites of
a chain (the ``matrix'' form).  These two examples are characterized by the
fact that the total Hamiltonian of the model is written as a sum of cell
Hamiltonians (which are not mutually commuting), and the exact ground-state
wave function of the total system is the eigenfunction having the lowest
energy of each cell Hamiltonian.

We have previously\cite{16} investigated an exactly solvable,
one-dimen\-sional, frustrated model with spin~1/2, whose properties by and
large are similar to those of the AKLT model.  The ground-state wave
function has a special recursion formula, and we have shown that it can be
reduced to matrix form.  It must be noted, however, that both the recursive
form and the matrix form are essentially one-dimen\-sional constructions
and cannot be extended directly to higher-dimen\-sional systems.  We cite
Ref.~\onlinecite{17} in this regard, where a method has been proposed for
constructing an exact wave function of the ground state for models with
spin 3/2 on a hexagonal lattice.  The same method is applicable to other
systems with spin~$d/2$.

In this paper we consider a class of models with spin~1/2 for which the
exact wave function of the ground state can be represented in an
alternative form.  In the one-dimen\-sional case this wave function reduces
to a wave function that we have found previously,\cite{16} but it admits
generalization to higher-dimen\-sional systems.  The present study is
devoted primarily to an analysis of the two-dimen\-sional model.

The article is organized as follows.  In Sec.~\ref{s2} we discuss the
method of construction of the exact wave function for a one-dimen\-sional
model with $s = 1/2$.  In Sec.~\ref{s3} we formulate an exactly solvable
two-dimen\-sional model.  In Sec.~\ref{s4} we investigate the properties of
this model with the aid of numerical calculations.  In Sec.~\ref{s5} we
discuss the possibility of generalizing our treatment to other types of
lattices.  The Appendix gives a proof of the nondegeneracy of the ground
state of the two-dimen\-sional model in the presence of cyclic boundary
conditions.

\section[]{One-dimensional model}\label{s2}

We have previously\cite{16} investigated a one-dimen\-sional, one-parameter
model containing two spins~1/2 in the unit cell and admitting exchange
interactions between nearest neighbor spins and spins separated by two
sites of the lattice.  The exact ground-state wave function of the cyclic
chain can be written in the form
%Eq(s). (1)
\begin{eqnarray}
\Psi _{0} = \text{Tr}\,\big [A(1,\, 2)\,A(3,\, 4)\,\ldots \,A(N - 1,\, N)
\big ],
\label{e1}
\end{eqnarray}
where $A(2i - 1,\, 2i)$ is a $2\times 2$ matrix associated with the {\it
i}th unit cell.

Below we write the wave function $\Psi _{0}$ in a form more suitable for
subsequent generalization to other types of lattices and give the general
form of the Hamiltonian for which $\Psi _{0}$ is an exact wave function of
the ground state.

We consider a chain of $N = 2M$ spins~1/2.  The wave function of this
system is described by the {\it N}th-rank spinor
%Eq(s). (2)
\begin{eqnarray}
\Psi = \Psi ^{\lambda \mu \nu \ldots \tau },
\label{e2}
\end{eqnarray}
where the indices $\lambda ,\, \mu ,\, \nu ,\ldots,\, \tau = 1,\, 2$
correspond to different projections of the spin~1/2.

We partition the system into pairs of nearest neighbor spins.  The wave
function can then be written as the product of {\it M} second-rank spinors
%Eq(s). (3)
\begin{eqnarray}
\Psi = \Psi ^{\lambda \mu }(1)\Psi ^{\nu \rho }(2)\ldots \Psi ^{\sigma \tau
}(M).
\label{e3}
\end{eqnarray}
We now form a scalar from Eq.~(\ref{e3}), simplifying the latter with
respect to index pairs:
%Eq(s). (4)
\begin{eqnarray}
\Psi _{s} = \Psi ^{\lambda }{}_{\nu }(1)\Psi ^{\nu }{}_{\kappa }(2)\ldots
\Psi ^{\sigma }{}_{\lambda }(M).
\label{e4}
\end{eqnarray}
Here subscripts correspond to the covariant components of the spinor, which
are related to the contravariant components (superscripts) through the
metric spinor
%Eq(s). (5-6)
\begin{eqnarray}
g_{\lambda \mu } = g^{\lambda \mu } = \left (
\begin{array}{cc}
0 & 1 \\
-1 & 0 \\
\end{array}
\right ). \label{e5} \\
\Psi _{\lambda } = g_{\lambda \mu }\Psi ^{\mu }, \qquad \Psi ^{\lambda } =
g^{\mu \lambda }\Psi _{\mu }.
\label{e6}
\end{eqnarray}

The scalar function (\ref{e4}) can thus be written in the form
%Eq(s). (7)
\begin{eqnarray}
\Psi _{s} = \Psi ^{\lambda \mu }(1)g_{\mu \nu }\Psi ^{\nu \rho }(2)g_{\rho
\kappa }\ldots \Psi ^{\sigma \tau}
(M)g_{\tau \lambda }.
\label{e7}
\end{eqnarray}
The scalar function $\Psi _{s}$ does not depend on the angle of rotation of
the coordinate system and, hence, corresponds to the singlet state.

The second-rank spinor describing the pair of spins~1/2 can be written in
the form
%Eq(s). (8)
\begin{eqnarray}
\Psi ^{\lambda \mu } = c_{t}\Psi _{t}^{\lambda \mu } + c_{s}\Psi
_{s}^{\lambda \mu },
\label{e8}
\end{eqnarray}
where $\Psi _{t}^{\lambda \mu }$ and $\Psi _{s}^{\lambda \mu }$ are
symmetric and antisymmetric second-rank spinors, respectively, and $c_{t}$
and $c_{s}$ are arbitrary constants.  We know that the symmetric
second-rank spinor describes a system with spin~1, so that the pair of
spins~1/2 in this case forms a triplet.  If $\Psi ^{\lambda \mu }$ is an
antisymmetric second-rank spinor reducible to a scalar multiplied by
$g_{\lambda \mu }$, the spin pair exists in the singlet state.
Consequently, the ratio of the constants $c_{t}$ and $c_{s}$ determines the
relative weights of the triplet and singlet components on the pair of spins
$s = 1/2$ and is a parameter of the model.  In particular, for $c_{s} = 0$
the wave function (\ref{e8}) contains only a triplet component, and for
$c_{t} = 0$ it contains only a singlet component.

In general, we can make the ratio of the constants $c_{s}/c_{t}$ different
in different pairs, but to preserve translational symmetry, we confine the
discussion to the case in which this ratio is the same in every pair.

We note that the wave function (\ref{e4}) has the matrix form (\ref{e1}),
the matrices $A(2i - 1,\, 2i)$ representing a mixed second-rank tensor:
%Eq(s). (9)
\begin{eqnarray}
A_{\lambda \nu }(1,2) = \Psi ^{\lambda }{}_{\nu }(1) = c_{t}\left (
\begin{array}{cc}
{\displaystyle \frac{1}{2}(\alpha _{1}\beta _{2} + \beta _{1}\alpha _{2})}
& \beta _{1}\beta _{2} \\
-\alpha _{1}\alpha _{2} & {\displaystyle -\frac{1}{2}(\alpha _{1}\beta _{2}
+ \beta _{1}\alpha _{2})} \\
\end{array} \right ) 
%\nonumber \\ [2mm]
- \frac{1}{2}(\alpha _{1}\beta _{2} - \beta _{1}\alpha _{2})\,I,
\label{e9}
\end{eqnarray}
where $\alpha _{i}$ and $\beta _{i}$ denote the up and down projections of
the spin ${\bf s}_{i}$, respectively, and {\it I} is the unit matrix.

We now choose a Hamiltonian {\it H} for which the wave function (\ref{e7})
is an exact ground-state wave function.  To do so, we consider the part of
the system (cell) consisting of two nearest neighbor spin pairs.  In the
wave function (\ref{e7}) the factor corresponding to the two spin pairs is
a second-rank spinor:
%Eq(s). (10)
\begin{eqnarray}
\Psi ^{\lambda \mu }(i)g_{\mu \nu }\Psi ^{\nu \rho }(i + 1).
\label{e10}
\end{eqnarray}
In the general case, therefore, only two of the six multiplets forming two
pairs of spin~1/2 --- one singlet and one triplet --- are present in the
wave function (\ref{e10}).  Inasmuch as four spins~1/2 form two singlets
and three triplets, the specific form of the singlet and triplet components
present in the wave function (\ref{e10}) depends on the ratio
$c_{s}/c_{t}$.  The cell Hamiltonian acting in the spin space of nearest
neighbor spin pairs can be written as the sum of the projectors onto the
four missing multiplets with arbitrary positive coefficients $\lambda
_{1},\, \lambda _{2},\, \lambda _{3},\, \lambda _{4}$:
%Eq(s). (11)
\begin{eqnarray}
H_{i,i + 1} = \sum \limits _{k = 1}^{4}\lambda _{k}P_{k}^{i,i + 1},
\label{e11}
\end{eqnarray}
where $P_{k}^{i,i + 1}$ is the projector onto the missing multiplets in the
corresponding cell Hamiltonian.

The wave function (\ref{e7}) is now an exact wave function of the ground
state of the cell Hamiltonian $H_{i,i + 1}$ with zero energy, because
%Eq(s). (12)
\begin{eqnarray}
H_{i,i + 1}|\Psi _{s}\rangle = 0,
\label{e12}
\end{eqnarray}
and $\lambda _{1},\, \lambda _{2},\, \lambda _{3},\, \lambda _{4}$ are the
excitation energies of the corresponding multiplets.

The total Hamiltonian of the entire system can be written as the sum of
mutually noncommuting cell Hamiltonians:
%Eq(s). (13)
\begin{eqnarray}
H = \sum \limits _{i = 1}^{N}H_{i,i + 1},
\label{e13}
\end{eqnarray}
and since each term $H_{i,i + 1}$ in (\ref{e13}) yields zero in its action
on $\Psi _{s}$, we have
%Eq(s). (14)
\begin{eqnarray}
H|\Psi _{s}\rangle = 0.
\label{e14}
\end{eqnarray}
The nondegeneracy of the ground state of this Hamiltonian has been
rigorously proved.\cite{16}

Since the specific form of the existing and missing multiplets in the wave
function (\ref{e7}) on every two nearest neighbor spin pairs depends on the
model parameter $c_{s}/c_{t}$, the projectors in (\ref{e11}) also depend on
$c_{s}/c_{t}$.  Each projector can be written in the form
%Eq(s). (15)
\begin{eqnarray}
P_{k}^{1,2} &=& J_{12}^{(k)}({\bf s}_{1}\cdot {\bf s}_{2} + {\bf
s}_{3}\cdot {\bf s}_{4}) + J_{13}^{(k)}({\bf s}_{1}\cdot {\bf s}_{3} + {\bf
s}_{2}\cdot {\bf s}_{4}) + J_{14}^{(k)}{\bf s}_{1}\cdot {\bf s}_{4}
\nonumber \\
&+& J_{23}^{(k)}{\bf s}_{2}\cdot {\bf s}_{3} + J_{1}^{(k)}({\bf s}_{1}\cdot
{\bf s}_{2})({\bf s}_{3}\cdot {\bf s}_{4}) + J_{2}^{(k)}({\bf s}_{1}\cdot
{\bf s}_{3})({\bf s}_{2}\cdot {\bf s}_{4}) \nonumber \\
&+& J_{3}^{(k)}({\bf s}_{1}\cdot {\bf s}_{4})({\bf s}_{2}\cdot {\bf s}_{3})
+ C^{(k)},
\label{e15}
\end{eqnarray}
and this representation is unique for a fixed value of the parameter
$c_{s}/c_{t}$.

Substituting the above expressions for the projectors into Eq.~(\ref{e11}),
we obtain the general form of the cell Hamiltonians $H_{i,i + 1}$.
Inasmuch as the Hamiltonians $H_{i,i + 1}$ have exactly the same form for
any {\it i}, it suffices here to give the expression for~$H_{1,2}$:
%Eq(s). (16)
\begin{eqnarray}
H_{1,2} &=& J_{12}({\bf s}_{1}\cdot {\bf s}_{2} + {\bf s}_{3}\cdot {\bf
s}_{4}) + J_{13}({\bf s}_{1}\cdot {\bf s}_{3} + {\bf s}_{2}\cdot {\bf
s}_{4}) + J_{14}{\bf s}_{1}\cdot {\bf s}_{4} + J_{23}{\bf s}_{2}\cdot {\bf
s}_{3} \nonumber \\
&+& J_{1}({\bf s}_{1}\cdot {\bf s}_{2})({\bf s}_{3}\cdot {\bf s}_{4}) +
J_{2}({\bf s}_{1}\cdot {\bf s}_{3})({\bf s}_{2}\cdot {\bf s}_{4}) +
J_{3}({\bf s}_{1}\cdot {\bf s}_{4})({\bf s}_{2}\cdot {\bf s}_{3}) + C,
\label{e16}
\end{eqnarray}
where all volume integrals depend on the model parameter and the spectrum
of excited states $J_{i} = J_{i}(c_{s}/c_{t},\, \lambda _{1},\, \lambda
_{2},\, \lambda _{3},\, \lambda _{4})$.  In particular, for $c_{s} = 0$,
choosing $\lambda _{2} = \lambda _{3} = \lambda _{4}$ and $\lambda
_{1}/\lambda _{2} = 3$, we obtain an expression for $H_{1,2}$ in the form
%Eq(s). (17)
\begin{eqnarray}
H_{1,2} = {\bf L}_{1}\cdot {\bf L}_{2} + \frac{1}{3}({\bf L}_{1}\cdot {\bf
L}_{2})^{2} + \frac{2}{3},
\label{e17}
\end{eqnarray}
where ${\bf L}_{1} = {\bf s}_{1} + {\bf s}_{2}$ and ${\bf L}_{2} = {\bf
s}_{3} + {\bf s}_{4}$.

The Hamiltonian (\ref{e17}) has the form of the AKLT Hamiltonian, a result
that is not too surprising, because for $c_{s} = 0$ two spins~1/2 in a pair
effectively form spin~1.  Note, however, that for $c_{s} = 0$ a set of
different forms of the Hamiltonian $H_{1,2}$ exists, corresponding to a
different choice of coefficients~$\lambda _{k}$.

In general, the Hamiltonian (\ref{e16}) contains both bilinear and
four-spin interactions.  The latter can be excluded by setting $J_{1} =
J_{2} = J_{3} = 0$ and solving these equations for $\lambda _{1},\, \lambda
_{2},\, \lambda _{3},\, \lambda _{4}$.  However, since the condition
$\lambda _{1},\, \lambda _{2},\, \lambda _{3},\, \lambda _{4} > 0$,
generally speaking, is not satisfied over the entire range of the parameter
$c_{s}/c_{t}$, the simplified Hamiltonian will also have a ground state
described by the wave function (\ref{e7}) only in the region where $\lambda
_{1},\, \lambda _{2},\, \lambda _{3},\, \lambda _{4}$ are positive.  The
nonzero exchange integrals $J_{12},\, J_{13},\, J_{14},\, J_{23}$ and the
constant {\it C} depend only on the parameter $c_{s}/c_{t}$.  The explicit
form of this dependence is given in Ref.~\onlinecite{16}, in which we have
also calculated the ground-state spin correlation function $\langle {\bf
s}_{i}{\bf s}_{j}\rangle $, which decays exponentially with correlation
length~$\sim 1$.

We emphasize that the spin correlation functions $\langle {\bf s}_{i}{\bf
s}_{j}\rangle $ do not depend on the choice of $\lambda _{1},\, \lambda
_{2},\,  \lambda _{3},\, \lambda _{4}$ for a fixed parameter $c_{s}/c_{t}$,
because the ground-state wave function of the four-parameter set of
Hamiltonians is the same.

\section[]{Two-dimensional model}\label{s3}

\begin{figure}[t]
\unitlength1cm
\begin{picture}(11,6)
\centerline{\psfig{file=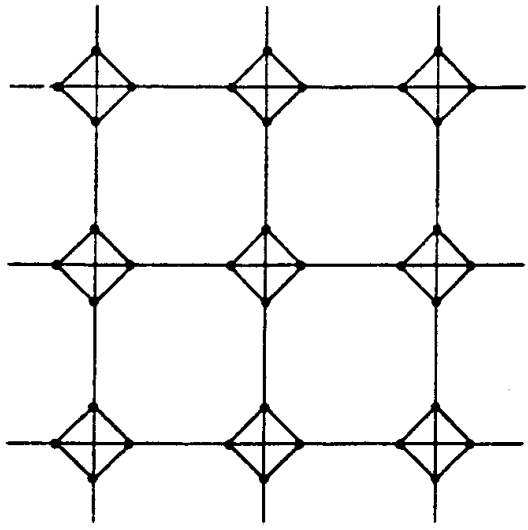,angle=0,width=8cm}}
\end{picture}
%\vspace{-2cm}
\caption[]{Two-dimensional lattice on which the spin model is
defined.\label{f1}}
\end{figure}

We consider an $M\times M$-site square lattice with cyclic boundary
conditions.  We replace each site of the lattice by a square
(Fig.~\ref{f1}) with spins $s = 1/2$ at its corners, making the total
number of spins equal to $4M^{2}$.  To avoid misunderstanding, however,
from now on we continue to refer to these squares as sites.  The wave
function of the system is described by the product of fourth-rank spinors
%Eq(s). (18)
\begin{eqnarray}
\Psi = \prod \limits _{{\bf n}}\Psi ^{\lambda _{{\bf n}}\mu _{{\bf n}}\nu
_{{\bf n}}\rho _{{\bf n}}}({\bf n}).
\label{e18}
\end{eqnarray}
By analogy with (\ref{e7}), from Eq.~(\ref{e18}) we form the scalar
%Eq(s). (19)
\begin{eqnarray}
\Psi = \prod \limits _{{\bf n}}\Psi ^{\lambda _{{\bf n}}\mu _{{\bf n}}\nu
_{{\bf n}}\rho _{{\bf n}}}({\bf n})g_{\nu _{{\bf n}} \lambda _{{\bf n} +
{\bf a}}}g_{\rho _{{\bf n}} \mu _{{\bf n} + {\bf b}}}.
\label{e19}
\end{eqnarray}
where {\bf a} and {\bf b} are unit vectors in the {\it x} and {\it y}
directions.

\begin{figure}[t]
\unitlength1cm
\begin{picture}(11,6)
\centerline{\psfig{file=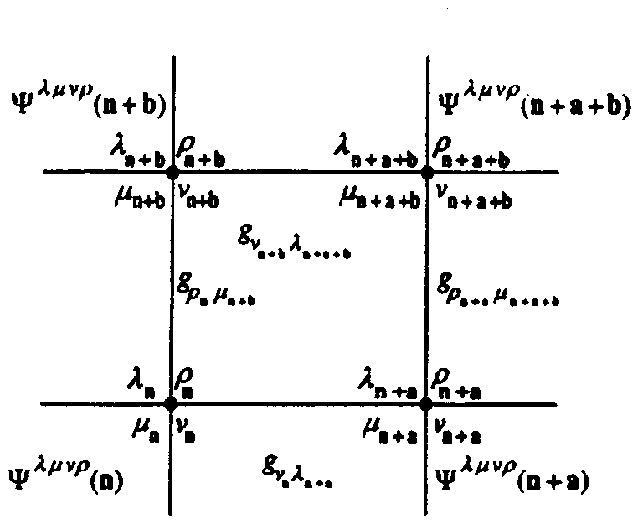,angle=0,width=8cm}}
\end{picture}
%\vspace{-2cm}
\caption[]{Graphical correspondence of the model wave function.  The
indices of the site spinors depend on the site index (not shown in the
figure).\label{f2}}
\end{figure}

The singlet wave function (\ref{e19}) is conveniently identified
graphically with a square lattice, each site corresponding to a fourth-rank
spinor $\Psi ^{\lambda \mu \nu \rho }$ (whose form is identical for all
sites), and each segment linking sites corresponds to a metric spinor
$g_{\lambda \mu }$ (Fig.~\ref{f2}).

To completely define the wave function (\ref{e19}), it is necessary to know
the form of the site spinor $\Psi ^{\lambda \mu \nu \rho }$.  For this
purpose we classify an arbitrary fourth-rank spinor, simplifying and
symmetrizing it with respect to different pairs of indices.  We have the
following types of spinors as a result:

1) a fourth-rank spinor $Q^{\lambda \mu \nu \rho }$ symmetric with respect
to all indices;

2) three linearly independent products of a symmetric and an antisymmetric
second-rank spinor: $\varphi ^{\lambda \mu }g_{\nu \rho }$, $\varphi
^{\lambda \nu }g_{\mu \rho }$, and $\varphi ^{\lambda \rho }g_{\mu \nu }$;

3) two linearly independent products of two metric spinors and a scalar
function: $g_{\lambda \mu }g_{\nu \rho }\chi $ and $g_{\lambda \nu }g_{\mu
\rho }\chi $.

According to this classification, any fourth-rank spinor can be written in
the form
%Eq(s). (20)
\begin{eqnarray}
\Psi ^{\lambda \mu \nu \rho } = c_{1}Q^{\lambda \mu \nu \rho } +
c_{2}\varphi _{1}^{\lambda \mu }g_{\nu \rho } + c_{3}\varphi _{2}^{\lambda
\nu }g_{\mu \rho } + c_{4}\varphi _{3}^{\lambda \rho}g_{\mu \nu } +
c_{5}g_{\lambda \mu }g_{\nu \rho }\chi _{1} + c_{6}g_{\lambda \nu }g_{\mu
\rho }\chi _{2}.
\label{e20}
\end{eqnarray}
We note, however, that because the system of four spins~1/2 contains one
quintet, three triplets, and two singlets, Eq.~(\ref{e20}) still does not
completely determine the form of $\Psi ^{\lambda \mu \nu \rho }$, and it is
necessary to determine the specific form of the spinors $\varphi
_{1}^{\lambda \mu }$, $\varphi _{2}^{\lambda \nu }$, and $\varphi
_{3}^{\lambda \rho }$ and the scalar functions $\chi _{1}$ and~$\chi _{2}$.

Each symmetric second-rank spinor $\varphi ^{\lambda \mu }$ describes a
triplet state of the system, representing a linear combination of the three
basis triplet functions $\varphi _{t1}^{\lambda \mu }$, $\varphi
_{t2}^{\lambda \mu }$, and $\varphi _{t3}^{\lambda \mu }$.  We can now
specify nine linearly independent spinors describing triplet states of four
spins $s = 1/2$:
%Eq(s). (21)
\begin{eqnarray}
\varphi _{t1}^{\lambda \mu }g_{\nu \rho },\,\,
\varphi _{t2}^{\lambda \mu }g_{\nu \rho },\,\,
\varphi _{t3}^{\lambda \mu }g_{\nu \rho }, \nonumber \\
\varphi _{t1}^{\lambda \nu }g_{\mu \rho },\,\,
\varphi _{t2}^{\lambda \nu }g_{\mu \rho },\,\,
\varphi _{t3}^{\lambda \nu }g_{\mu \rho }, \label{e21} \\
\varphi _{t1}^{\lambda \rho }g_{\mu \nu },\,\,
\varphi _{t2}^{\lambda \rho }g_{\mu \nu },\,\,
\varphi _{t3}^{\lambda \rho }g_{\mu \nu }. \nonumber
\end{eqnarray}
The products of two metric spinors and a scalar function $g_{\lambda \mu
}g_{\nu \rho }\chi _{1}$ and $g_{\lambda \nu }g_{\mu \rho }\chi _{2}$
describe singlets states of four spins $s = 1/2$.  Since there are two
independent singlet functions $\chi _{s1}$ and $\chi _{s2}$, we have four
linearly independent scalars describing singlet states of four spins $s =
1/2$:
%Eq(s). (22)
\begin{eqnarray}
\begin{array}{c}
g_{\lambda \mu }g_{\nu \rho }\chi _{s1},\, g_{\lambda \mu }g_{\nu \rho
}\chi _{s2}, \\
g_{\lambda \nu }g_{\mu \rho }\chi _{s1},\, g_{\lambda \nu }g_{\mu \rho
}\chi _{s2}. \\
\end{array}
\label{e22}
\end{eqnarray}

As a result, the specific form of the fourth-rank spinor $\Psi ^{\lambda
\mu \nu \rho }$ [and, hence, the wave function (\ref{e19})] describing the
system of four spins $s = 1/2$ is governed by $1 + 9 + 4 = 14$ quantities,
which are parameters of the model.

We now choose a Hamiltonian {\it H} for which the wave function (\ref{e19})
is an exact ground-state wave function.  As in the one-dimen\-sional case,
we seek the required Hamiltonian in the form of a sum of cell Hamiltonians
acting in the space of two nearest neighbor spin quartets:
%Eq(s). (23)
\begin{eqnarray}
H = \sum \limits _{{\bf n}}H_{{\bf n},{\bf n} + {\bf a}} +
    \sum \limits _{{\bf n}}H_{{\bf n},{\bf n} + {\bf b}}.
\label{e23}
\end{eqnarray}

\begin{figure}[t]
\unitlength1cm
\begin{picture}(11,6)
\centerline{\psfig{file=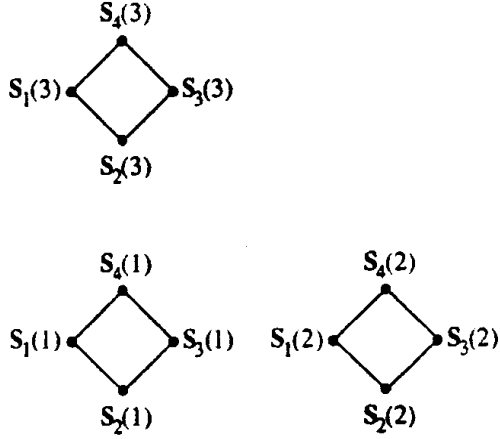,angle=0,width=8cm}}
\end{picture}
%\vspace{-2cm}
\caption[]{Lattice sites associated with interactions $H_{1,2}$
and~$H_{1,3}$.\label{f3}}
\end{figure}

The first term in Eq.~(\ref{e23}) is the sum of the cell Hamiltonians in
the horizontal direction, and the second term is the same for the vertical.
The cell Hamiltonians along each direction have the same form, but the
``horizontal'' and ``vertical'' Hamiltonians differ in general.  In the
ensuing discussion, therefore, we consider only the Hamiltonians $H_{1,2}$
and $H_{1,3}$ (Fig.~\ref{f3}), which describe interactions of ``sites'' in
the {\it x} and {\it y} directions, respectively.

For the wave function (\ref{e19}) to be an exact eigenfunction of the
Hamiltonian {\it H}, it is sufficient that the sixth-rank spinors
%Eq(s). (24-25)
\begin{eqnarray}
&{}&\Psi ^{\lambda _{1}\mu _{1}\nu _{1}\rho _{1}}(1)
\Psi ^{\lambda _{2}\mu _{2}\nu _{2}\rho _{2}}(2)
g_{\nu _{1}\lambda _{2}},
\label{e24} \\
&{}&\Psi ^{\lambda _{1}\mu _{1}\nu _{1}\rho _{1}}(1)
\Psi ^{\lambda _{3}\mu _{3}\nu _{3}\rho _{3}}(3)
g_{\rho _{1}\mu _{3}},
\label{e25}
\end{eqnarray}
be eigenfunctions of the corresponding cell Hamiltonians $H_{1,2}$
and~$H_{1,3}$.

In general, when the site spinor $\Psi ^{\lambda \mu \nu \rho }$ is not
symmetric with respect to any indices, the possible states of two quartets
of spins $s = 1/2$ consist of 70 multiplets.  A wave function represented
by a sixth-rank spinor contains only 20 of them.  Accordingly, the cell
Hamiltonians $H_{1,2}$ and $H_{1,3}$ can be represented by the sum of
projectors onto the 50 missing multiplets:
%Eq(s). (26)
\begin{eqnarray}
H_{1,2} = \sum \limits _{k = 1}^{50}\lambda _{k}P_{k}^{1,2}, \nonumber \\
[2mm]
H_{1,3} = \sum \limits _{k = 1}^{50}\mu _{k}P_{k}^{1,3},
\label{e26}
\end{eqnarray}
where the positive constants $\lambda _{k}$ and $\mu _{k}$ are the
excitation energies of $H_{1,2}$ and $H_{1,3}$, and the specific form of
the projectors depends on 14 model parameters.

Inasmuch as
%Eq(s). (27)
\begin{eqnarray}
H_{{\bf n},{\bf n} + {\bf a}}|\Psi _{s}\rangle = 0, \quad
H_{{\bf n},{\bf n} + {\bf b}}|\Psi _{s}\rangle = 0,
\label{e27}
\end{eqnarray}
for the total Hamiltonian (\ref{e23}) we have the expression
%Eq(s). (28)
\begin{eqnarray}
H|\Psi _{s}\rangle = 0.
\label{e28}
\end{eqnarray}

Consequently, $\Psi _{s}$ is the ground-state wave function of the total
Hamiltonian {\it H}, because it is a sum of nonnegative definite cell
Hamiltonians.  Also, it can be rigorously proved (see the Appendix) that
the ground state of {\it H} is nondegenerate.

As mentioned above, the specific form of the projectors depends on 14 model
parameters, and in general the cell Hamiltonians (\ref{e26}), expressed in
terms of scalar products of the type ${\bf s}_{i}\cdot {\bf s}_{j}$, $({\bf
s}_{i}\cdot {\bf s}_{j})({\bf s}_{k}\cdot {\bf s}_{l})$, etc., have an
extremely cumbersome form.  We therefore consider a few special cases.

When the site spinor $\Psi ^{\lambda \mu \nu \rho }$ is a symmetric
fourth-rank spinor $Q^{\lambda \mu \nu \rho }$ (corresponding to the
two-dimen\-sional AKLT model\cite{12}), only the quintet component out of the
six multiplets on each spin quartet is present in the wave function
(\ref{e19}).  The sixth-rank spinors (\ref{e24}) and (\ref{e25}) are
symmetric with respect to two triplets of indices and, hence, contain four
multiplets with $S = 0,\, 1,\, 2,\, 3$ formed from two quintets.
Consequently, the cell Hamiltonian ($H_{1,2}$ and $H_{1,3}$ coincide in
this case) has the form
%Eq(s). (29)
\begin{eqnarray}
H_{1,2} = \sum \limits _{k = 1}^{66}\lambda _{k}P_{k}^{1,2}.
\label{e29}
\end{eqnarray}
If we set $\lambda _{k} = 1$ ($k = 1,\, 66$), we can write Eq.~(\ref{e29})
in the form
%Eq(s). (30)
\begin{eqnarray}
H_{1,2} = P_{4}({\bf S}_{1} + {\bf S}_{2}) + \big [1 - P_{2}({\bf
S}_{1})P_{2}({\bf S}_{2})\big ],
\label{e30}
\end{eqnarray}
where ${\bf S}_{i}$ is the total spin of the quartet of spins $s = 1/2$ on
the {\it i}th site, ${\bf S}_{i} = {\bf s}_{1}(i) + {\bf s}_{2}(i) + {\bf
s}_{3}(i) + {\bf s}_{4}(i)$, and $P_{l}({\bf S})$ is the projector onto
the state with spin $S = l$.

If the four spins $s = 1/2$ at each site are replaced by a single spin $S =
2$ and if the wave function (\ref{e19}) is treated as a wave function
describing a system of $M^{2}$ spins $S = 2$, the second term in the
Hamiltonian (\ref{e30}) vanishes, and we arrive at the Hamiltonian of the
two-dimen\-sional AKLT model:
%Eq(s). (31)
\begin{eqnarray}
H_{1,2} = P_{4}({\bf S}_{1} + {\bf S}_{2}) = \frac{1}{28}{\bf S}_{1}\cdot
{\bf S}_{2} + \frac{1}{40}({\bf S}_{1}\cdot {\bf S}_{2})^{2} +
\frac{1}{180}({\bf S}_{1}\cdot {\bf S}_{2})^{3} + \frac{1}{2520}({\bf
S}_{1}\cdot {\bf S}_{2})^{4}.
\label{e31}
\end{eqnarray}

Another interesting special case is encountered when the system decomposes
into independent one-dimen\-sional chains.  This happens if the site spinor
$\Psi ^{\lambda \mu \nu \rho }$ reduces to a product of two second-rank
spinors, each describing two spins~1/2.  For example,
%Eq(s). (32)
\begin{eqnarray}
\Psi ^{\lambda \mu \nu \rho }(s_{1},\, s_{2},\, s_{3},\, s_{4}) = \varphi
^{\lambda \nu }(s_{1},\, s_{3})\varphi ^{\mu \rho }(s_{2},\, s_{4}).
\label{e32}
\end{eqnarray}
In this case the Hamiltonians $H_{1,2}$ and $H_{1,3}$ contain interactions
of four rather than eight spins~1/2 and have the form~(\ref{e16}).

\begin{figure}[t]
\unitlength1cm
\begin{picture}(11,6)
\centerline{\psfig{file=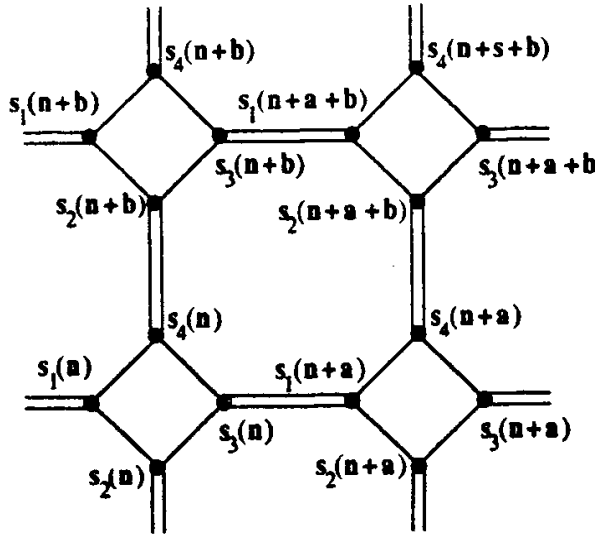,angle=0,width=8cm}}
\end{picture}
%\vspace{-2cm}
\caption[]{Pattern of independent singlet pairs (double lines).\label{f4}}
\end{figure}

The simplest case is when the site spinor $\Psi ^{\lambda \mu \nu \rho }$
is a product of four first-rank spinors:
%Eq(s). (33)
\begin{eqnarray}
\Psi ^{\lambda \mu \nu \rho }(s_{1},\, s_{2},\, s_{3},\, s_{4}) = \varphi
^{\lambda }(s_{1})\varphi ^{\mu }(s_{2})\varphi ^{\nu }(s_{3})\varphi
^{\rho }(s_{4}).
\label{e33}
\end{eqnarray}
Now the system decomposes into independent singlet pairs (Fig.~\ref{f4}),
and the total Hamiltonian of the system has the form
%Eq(s). (34)
\begin{eqnarray}
H = \sum \limits _{i,j}\left ({\bf s}_{i}\cdot {\bf s}_{j} +
\frac{3}{4}\right ),
\label{e34}
\end{eqnarray}
where ${\bf s}_{i}$ and ${\bf s}_{j}$ are the spins forming the singlet
pairs.

\section[]{Spin correlation functions in the ground state}\label{s4}

We now look at the problem of calculating the norm and the correlation
function of the model described by the wave function (\ref{e19}).  The
expression for the norm of the wave function $G = \langle \Psi _{s}|\Psi
_{s}\rangle $ has the form
%Eq(s). (35)
\begin{eqnarray}
G &=& \prod \limits _{{\bf n}}\left \langle
\Psi ^{\lambda '_{{\bf n}}\mu '_{{\bf n}}\nu '_{{\bf n}}\rho ' _{{\bf
n}}}({\bf n})\Big |
\Psi ^{\lambda _{{\bf n}}\mu _{{\bf n}}\nu _{{\bf n}}\rho
_{{\bf n}}}({\bf n})\right \rangle
g_{\nu _{{\bf n}}\lambda _{{\bf n} + {\bf a}}}
g_{\rho _{{\bf n}}\mu _{{\bf n} + {\bf b}}}
g_{\nu '_{{\bf n}}\lambda '_{{\bf n} + {\bf a}}}
g_{\rho '_{{\bf n}}\mu '_{{\bf n} + {\bf b}}} \nonumber \\ [3mm]
&=& \prod \limits _{{\bf n}}
R_{\lambda _{{\bf n}}\mu _{{\bf n}}\lambda _{{\bf n} + {\bf a}}\mu _{{\bf
n} + {\bf b}}}
^{\lambda '_{{\bf n}}\mu '_{{\bf n}}\lambda '_{{\bf n} + {\bf a}}\mu
'_{{\bf n} + {\bf b}}} =
\prod \limits _{{\bf n}}
R_{\alpha _{{\bf n}}\beta _{{\bf n}}\alpha _{{\bf n} + {\bf a}}\beta _{{\bf
n} + {\bf b}}},
\qquad \alpha _{i},\, \beta _{i} = \{1,\, 2,\, 3,\, 4\},
\label{e35}
\end{eqnarray}
where $R_{\alpha _{}\beta _{}\alpha _{}\beta _{}}$ is a $4\times 4\times
4\times 4$ matrix.

According to the selection rules for the projection of the total spin
$S^{z}$, only 70 of the 256 elements in the expression $\Bigl \langle \Psi
^{\lambda '_{}\mu '_{}\nu '_{}\rho _{}}({\bf n})\Big |\Psi ^{\lambda _{}\mu
_{}\nu _{}\rho _{}}({\bf n})\Big \rangle $ are nonvanishing.  Consequently,
the matrix {\it R} also contains at most 70 elements.  If we regard the
elements of {\it R} as Boltzmann vertex weights, the problem of calculating
the norm reduces to the classical 70-vertex model.

Since the exact solution for the 70-vertex model is unknown, numerical
methods must be used to calculate the norm and the expected values.

To calculate the above-indicated expected values, we carry out Monte Carlo
calculations on $20\times 20$-site lattices.  As mentioned, the
ground-state wave function of the model depends on 14 parameters and, of
course, cannot possibly be analyzed completely.  We confine the numerical
calculations to the case in which the spinor $\Psi ^{\lambda \mu \nu \rho
}$ depends on one parameter~$\alpha $:
%Eq(s). (36)
\begin{eqnarray}
\Psi ^{\lambda \mu \nu \rho } = \cos \alpha \cdot Q^{\lambda \mu \nu \rho }
+ \sin \alpha \cdot \Big (A^{\lambda \mu \nu \rho } - Q^{\lambda \mu \nu
\rho }\Big ),
\label{e36}
\end{eqnarray}
where $\alpha \in \big [-\pi /2;\,\, \pi /2\big ]$, the spinor $Q^{\lambda
\mu \nu \rho }$ is symmetric with respect to all indices, and
%Eq(s). (37)
\begin{eqnarray}
A^{\lambda \mu \nu \rho } =
\varphi ^{\lambda }(s_{1})\varphi ^{\mu }(s_{2})
\varphi ^{\nu }(s_{3})\varphi ^{\rho }(s_{4}).
\label{e37}
\end{eqnarray}

In this case we have a one-parameter model with two well-known limiting
cases.  One corresponds to $\alpha  = \pi /4$, for which $\Psi ^{\lambda
\mu \nu \rho } = A^{\lambda \mu \nu \rho }$, and the system decomposes into
independent singlet pairs (Fig.~\ref{f4}); the other limiting case
corresponds to $\alpha = 0$ (our model reduces to the two-dimen\-sional
AKLT model in this case, the spins at each site forming a quintet).

In the given model there are four spins $s = 1/2$ at each site, and the
enumeration of each spin is determined by the order number of the lattice
site to which it belongs and by its own number at this site.  The spin
correlation function therefore has the form
%Eq(s). (38)
\begin{eqnarray}
f_{ij}({\bf r}) = \big \langle {\bf s}_{i}({\bf n})\cdot {\bf s}_{j}({\bf
n} + {\bf r})\big \rangle .
\label{e38}
\end{eqnarray}

In determining the spin structure of the ground state, however, it is more
practical to consider the more straightforward quantity~$F({\bf r})$:
%Eq(s). (39)
\begin{eqnarray}
F({\bf r}) = \sum \limits _{i,j = 1}^{4}\big \langle {\bf s}_{i}({\bf
n})\cdot {\bf s}_{j}({\bf n} + {\bf r})\big \rangle =
\big \langle {\bf S}({\bf n})\cdot {\bf S}({\bf n} + {\bf r})\big \rangle.
\label{e39}
\end{eqnarray}

The function $F({\bf r})$ is left unchanged by a change of sign of $\alpha
$.  This invariance is attributable to the fact that the spinor
$(A^{\lambda \mu \nu \rho } - Q^{\lambda \mu \nu \rho }$ does not contain a
quintet component, so that all the functions of this spinor are orthogonal
to all functions of the symmetric spinor
%Eq(s). (40)
\begin{eqnarray}
\left \langle Q^{\lambda '\mu '\nu '\rho '}\Big |
\Big (A^{\lambda \mu \nu \rho } - Q^{\lambda \mu \nu \rho }\Big )
\right \rangle  = 0.
\label{e40}
\end{eqnarray}
for all $\lambda ,\, \mu ,\, \nu ,\, \rho $ and $\lambda ',\, \mu ',\, \nu
',\, \rho '$.

In addition, since the total spin operator {\bf S} at a site commutes with
${\bf S}^{2} = \sum \limits _{i,j = 1}^{4}{\bf s}_{i}\cdot {\bf s}_{j}$, we
then have
%Eq(s). (41)
\begin{eqnarray}
\left \langle Q^{\lambda '\mu '\nu '\rho '}\Big |
\sum \limits _{i = 1}^{4}{\bf s}_{i}\Big |
\Big (A^{\lambda \mu \nu \rho } - Q^{\lambda \mu \nu \rho }\Big )
\right \rangle  = 0
\label{e41}
\end{eqnarray}

\begin{figure}[t]
\unitlength1cm
\begin{picture}(11,6)
\centerline{\psfig{file=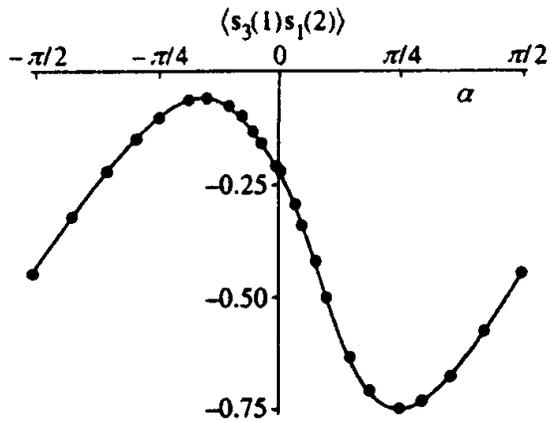,angle=0,width=8cm}}
\end{picture}
%\vspace{-2cm}
\caption[]{Dependence of the spin correlation function $\langle {\bf
s}_{3}(1){\bf s}_{1}(2)\rangle $ on the parameter~$\alpha $.\label{f5}}
\end{figure}

It follows from Eqs.~(\ref{e35}), (\ref{e40}), and (\ref{e41}) that $\sin
\alpha $ and $\cos \alpha $ enter into the norm and into the expected value
$\langle \Psi |{\bf S}({\bf n})\cdot {\bf S}({\bf n} + {\bf r})|\Psi
\rangle $ only in even powers, so that $F({\bf r})$ is invariant against a
change of sign of $\alpha $.  We note, however, that only the total
correlation function, and not $f_{ij}({\bf r})$, possesses symmetry under a
change of sign of $\alpha $.  This assertion is evident, for example, in
Fig.~\ref{f5}, which shows the dependence of $f_{31}({\bf a})$ on $\alpha $
as an illustration.

\begin{figure}[t]
\unitlength1cm
\begin{picture}(11,6)
\centerline{\psfig{file=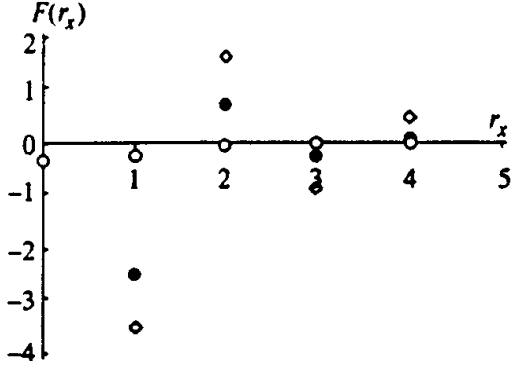,angle=0,width=8cm}}
\end{picture}
%\vspace{-2cm}
\caption[]{Dependence of the spin correlation function $F(r_{z})$ on the
distance along the {\it x}~axis for various values of the parameter~$\alpha
$: ($\diamond $)~$\alpha = 0$; ($\bullet $)~$\alpha = \pi/10$; ($\circ
$)~$\alpha = \pm \pi /2$.\label{f6}}
\end{figure}

\begin{figure}[t]
\unitlength1cm
\begin{picture}(11,6)
\centerline{\psfig{file=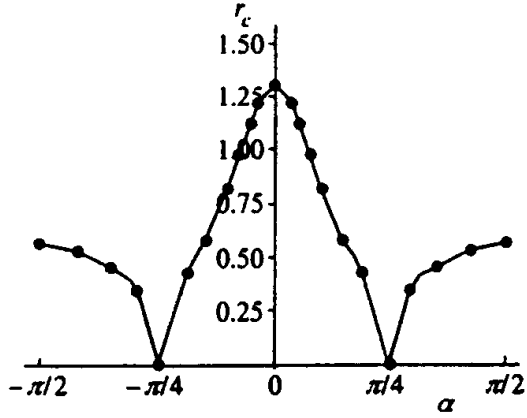,angle=0,width=8cm}}
\end{picture}
%\vspace{-2cm}
\caption[]{Dependence of the correlation length on the parameter~$\alpha
$.\label{f7}}
\end{figure}

\begin{figure}[t]
\unitlength1cm
\begin{picture}(11,6)
\centerline{\psfig{file=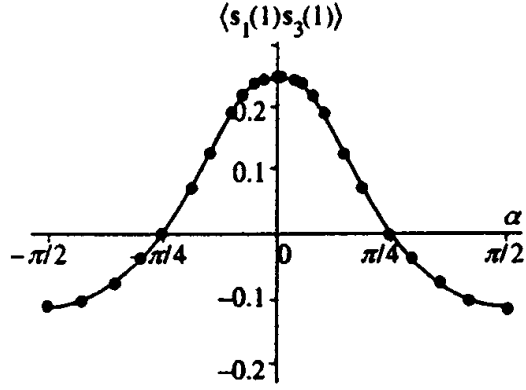,angle=0,width=8cm}}
\end{picture}
%\vspace{-2cm}
\caption[]{Dependence of the spin correlation function at one site on the
parameter~$\alpha $.\label{f8}}
\end{figure}

Figure~\ref{f6} shows plots of $F({\bf r})$ for certain values of the
parameter $\alpha $.  In every case it is found that the correlation
function decays exponentially as {\bf r} increases, differing from the
one-dimen\-sional model in that the preexponential factor also depends on
{\bf r}.  Figure~\ref{f7} shows the dependence of the correlation length
$r_{c}$ on the parameter $\alpha $.  The correlation length is a maximum at
the point $\alpha = 0$ (two-dimen\-sional AKLT model), decreases as $\alpha
$ increases, and at $\alpha = \pi /4$, when the system decomposes into
independent singlet pairs (Fig.~\ref{f4}), it is equal to zero.  With a
further increase in $\alpha $ the correlation length increases and attains
a second maximum at $\alpha = \pi /2$.  Like the correlation function
$F({\bf r})$, the function $r_{c}(\alpha )$ is symmetric with respect to
$\alpha $.  It is evident from Fig.~\ref{f7} that the parameter $\alpha $
has two rangess corresponding to states with different symmetries.  In the
range $|\alpha | < \pi /4$ the correlation function $F({\bf r})$ exhibits
antiferromagnetic behavior:
%Eq(s). (42)
\begin{eqnarray}
F({\bf r}) \propto (-1)^{r_{x} + r_{y}}e^{-|{\bf r}|/r_{c}},
\label{e42}
\end{eqnarray}
whereas the spins at one site are coupled ferromagnetically, $\langle {\bf
s}_{i}({\bf n})\cdot {\bf s}_{j}({\bf n})\rangle > 0$.  On the other hand,
in the range $\pi /4 < |\alpha | < \pi /2$ the correlation function $F({\bf
r})$ is always negative:
%Eq(s). (43)
\begin{eqnarray}
F({\bf r}) \propto -e^{-|{\bf r}|/r_{c}}
\label{e43}
\end{eqnarray}
and all the correlation functions at one site are also negative
(Fig.~\ref{f8}).

These ranges have two end points in common, $\alpha  = \pm \pi /4$, where
$r_{c} = 0$.  Whereas $\alpha = \pi /4$ corresponds to the trivial
partition of the system into independent singlet pairs, the case $\alpha
= -\pi /4$ is more interesting.

In this case we have
%Eq(s). (44)
\begin{eqnarray}
\Psi ^{\lambda \mu \nu \rho } = 2Q^{\lambda \mu \nu \rho } -A^{\lambda \mu
\nu \rho } ,
\label{e44}
\end{eqnarray}
and the matrix $\Big \langle \Psi ^{\lambda '\mu '\nu '\rho '}\Big |
\Psi ^{\lambda \mu \nu \rho }\Big \rangle $,
which enters into the equation for the norm
(\ref{e35}) and the expected values, is transformed to
%Eq(s). (45)
\begin{eqnarray}
\left \langle \Psi ^{\lambda '\mu '\nu '\rho '}\Big |
\Psi ^{\lambda \mu \nu \rho }\right \rangle &=&
4\left \langle Q^{\lambda '\mu '\nu '\rho '}\Big |
Q^{\lambda \mu \nu \rho }\right \rangle -
2\left \langle A^{\lambda '\mu '\nu '\rho '}\Big |
Q^{\lambda \mu \nu \rho }\right \rangle \nonumber \\ [2mm]
&-& 2\left \langle Q^{\lambda '\mu '\nu '\rho '}\Big |
A^{\lambda \mu \nu \rho }\right \rangle +
\left \langle A^{\lambda '\mu '\nu '\rho '}\Big |
A^{\lambda \mu \nu \rho }\right \rangle .
\label{e45}
\end{eqnarray}

The symmetry of the spinor $Q^{\lambda \mu \nu \rho }$ with respect to all
the indices leads to the relation
%Eq(s). (46)
\begin{eqnarray}
\left \langle Q^{\lambda '\mu '\nu '\rho '}\Big |
Q^{\lambda \mu \nu \rho }\right \rangle =
\left \langle A^{\lambda '\mu '\nu '\rho '}\Big |
Q^{\lambda \mu \nu \rho }\right \rangle =
\left \langle Q^{\lambda '\mu '\nu '\rho '}\Big |
A^{\lambda \mu \nu \rho }\right \rangle .
\label{e46}
\end{eqnarray}

Equation (\ref{e45}) therefore acquires the form
%Eq(s). (47)
\begin{eqnarray}
\left \langle \Psi ^{\lambda '\mu '\nu '\rho '}\Big |
\Psi ^{\lambda \mu \nu \rho }\right \rangle =
\left \langle A^{\lambda '\mu '\nu '\rho '}\Big |
A^{\lambda \mu \nu \rho }\right \rangle =
\delta _{\lambda \lambda '}\delta _{\mu \mu '}\delta _{\nu \nu '}\delta
_{\rho \rho '}.
\label{e47}
\end{eqnarray}

>From the equation for the norm (\ref{e35}) we then have
%Eq(s). (48)
\begin{eqnarray}
G &=& \prod \limits _{{\bf n}}\left \langle
\Psi ^{\lambda '_{{\bf n}}\mu '_{{\bf n}}\nu '_{{\bf n}}\rho ' _{{\bf
n}}}({\bf n})\Big |
\Psi ^{\lambda _{{\bf n}}\mu _{{\bf n}}\nu _{{\bf n}}\rho
_{{\bf n}}}({\bf n})\right \rangle
g_{\nu _{{\bf n}}\lambda _{{\bf n} + {\bf a}}}
g_{\rho _{{\bf n}}\mu _{{\bf n} + {\bf b}}}
g_{\nu '_{{\bf n}}\lambda '_{{\bf n} + {\bf a}}}
g_{\rho '_{{\bf n}}\mu '_{{\bf n} + {\bf b}}} \nonumber \\ [3mm]
&=& \prod \limits _{{\bf n}}
\delta _{\lambda _{{\bf n}}\lambda '_{{\bf n}}}\delta _{\mu _{{\bf n}}\mu
'_{{\bf n}}}
\delta _{\lambda _{{\bf n} + {\bf a}}\lambda '_{{\bf n} + {\bf a}}}\delta
_{\mu _{{\bf n} + {\bf b}}\mu '_{{\bf n} + {\bf b}}} = 2^{2M^{2}}\,.
\label{e48}
\end{eqnarray}
The latter equation has been derived with allowance for the relation
$\delta _{\nu \nu '}g_{\nu \lambda }g_{\nu '\lambda '} = \delta _{\lambda
\lambda '}$.

We now calculate the expected value $\langle \Psi |{\bf s}_{i}({\bf
n})\cdot {\bf s}_{j}({\bf n} + {\bf r})|\Psi \rangle $.  If sites {\bf n}
and ${\bf n} + {\bf r}$ are not nearest neighbors, $\langle \Psi |{\bf
s}_{i}({\bf n})\cdot {\bf s}_{j}({\bf n} + {\bf r})|\Psi \rangle $
decomposes into the product of the expected values
%Eq(s). (49)
\begin{eqnarray}
\big \langle \Psi \big |{\bf s}_{i}({\bf n})\cdot {\bf s}_{j}({\bf n} +
{\bf r})\big |\Psi \big \rangle &=&
2^{2M^{2} - 8}\left \langle \Psi ^{\lambda '\mu '\nu '\rho '}({\bf n})\bigg
|{\bf s}_{i}({\bf n})\Big |\Psi ^{\lambda \mu \nu \rho }({\bf n})\right
\rangle
\nonumber \\ [2mm]
&\times &
\left \langle \Psi ^{\lambda '''\mu '''\nu '''\rho '''}({\bf n} + {\bf
r})\Big |{\bf s}_{j}({\bf n} + {\bf r})\Big |\Psi ^{\lambda ''\mu ''\nu
''\rho ''}({\bf n} + {\bf r})\right \rangle \nonumber \\ [2mm]
&\times &
\delta _{\lambda \lambda '}\delta _{\mu \mu '}\delta _{\nu \nu '}\delta
_{\rho \rho '}
\delta _{\lambda ''\lambda '''}\delta _{\mu ''\mu '''}\delta _{\nu ''\nu
'''}\delta _{\rho ''\rho '''} = 0.
\label{e49}
\end{eqnarray}
Consequently, for $\alpha = -\pi /4$ all the correlation functions at
non-nearest neighbor sites are equal to zero.  But if sites {\bf n} and
${\bf n} + {\bf r}$ are nearest neighbors, the corresponding correlation
function assumes the form
%Eq(s). (50)
\begin{eqnarray}
\big \langle \Psi \big |{\bf s}_{i}(1)\cdot {\bf s}_{j}(2)
\big |\Psi \big \rangle &=&
2^{2M^{2} - 7}\left \langle \Psi ^{\lambda '\mu '\nu '\rho '}(1)\Big |{\bf
s}_{i}(1)\Big |\Psi ^{\lambda \mu \nu \rho }({\bf n})\right \rangle
\nonumber \\ [2mm]
&\times &
\left \langle \Psi ^{\lambda '''\mu '''\nu '''\rho '''}(2)\Big |{\bf
s}_{j}(2)\Big |\Psi ^{\lambda ''\mu ''\nu ''\rho ''}(2)\right \rangle
\nonumber \\ [2mm]
&\times &
g_{\nu \lambda ''}g_{\nu '\lambda '''}\delta _{\lambda \lambda '}
\delta _{\mu \mu '}\delta _{\rho \rho '}\delta _{\mu ''\mu '''}
\delta _{\nu ''\nu '''}\delta _{\rho ''\rho '''}.
\label{e50}
\end{eqnarray}
The exact calculation of the latter expression yields the following results
(Fig.~\ref{f3}):
%Eq(s). (51)
\begin{eqnarray}
\big \langle {\bf s}_{i}(1)\cdot {\bf s}_{j}(2)\big \rangle &=& -
\frac{25}{768}, \quad i = 1,\, 2,\, 4, \quad j = 2,\, 3,\, 4, \nonumber \\
[2mm]
\big \langle {\bf s}_{i}(1)\cdot {\bf s}_{1}(2)\big \rangle &=&
\big \langle {\bf s}_{3}(1)\cdot {\bf s}_{j}(2)\big \rangle =
-\frac{15}{256}, \label{e51} \\ [2mm]
\big \langle {\bf s}_{3}(1)\cdot {\bf s}_{1}(2)\big \rangle &=&
-\frac{27}{256}. \nonumber
\end{eqnarray}

It follows from Eqs.~(\ref{e51}) that
\begin{eqnarray*}
\sum \limits _{i,j = 1}^{4}\big \langle {\bf s}_{i}(1)\cdot {\bf
s}_{j}(2)\big \rangle = -\frac{3}{4},
\end{eqnarray*}
as in the case of independent singlets ($\alpha = \pi /4$).  It can also be
shown that all the correlations functions at one site are equal to zero.

to write the cell Hamiltonian $H_{1,2}$ in explicit form for $\alpha = -\pi
/4$, we introduce the notation
%Eq(s). (52)
\begin{eqnarray}
&{}&\left \{
\begin{array}{c}
{\bf l}_{1} = {\bf s}_{1}(1) + {\bf s}_{2}(1) + {\bf s}_{4}(1), \\
{\bf l}_{2} = {\bf s}_{2}(2) + {\bf s}_{3}(2) + {\bf s}_{4}(2), \\
\end{array} \right . \qquad
\left \{
\begin{array}{c}
{\bf s}_{1} = {\bf s}_{3}(1), \\
{\bf s}_{2} = {\bf s}_{1}(2), \\
\end{array} \right . \nonumber \\ [-1.5mm]
{} \label{e52} \\ [-1.5mm]
&{}&\left \{
\begin{array}{c}
{\bf h}_{1} = {\bf l}_{1}\cdot {\bf s}_{1} + {\bf l}_{2}\cdot {\bf s}_{2},
\\
{\bf h}_{2} = {\bf l}_{1}\cdot {\bf s}_{2} + {\bf l}_{2}\cdot {\bf s}_{1}.
\\
\end{array} \right .
\nonumber
\end{eqnarray}

Accordingly, choosing $\lambda _{k} = 1$ ($k = 1,\, 50$), we can write the
cell Hamiltonian $H_{1,2}$ in Eq.~(\ref{e26}) in the form
%Eq(s). (53)
\begin{eqnarray}
H_{1,2} &=& P_{1/2}({\bf l}_{1})P_{1/2}({\bf l}_{2})P_{1}({\bf s}_{1} +
{\bf s}_{2}) + P_{3/2}({\bf l}_{1})P_{3/2}({\bf l}_{2})h_{3} \nonumber \\
&+& P_{3/2}({\bf l}_{1})P_{1/2}({\bf l}_{2})h_{4} +
P_{1/2}({\bf l}_{1})P_{3/2}({\bf l}_{2})h_{5},
\label{e53}
\end{eqnarray}
where
%Eq(s). (54)
\begin{eqnarray}
\begin{array}{l}
{\displaystyle h_{3} = \frac{207}{256} + \frac{49}{64}{\bf s}_{1}\cdot {\bf
s}_{2} + \frac{3}{64}{\bf l}_{1}\cdot {\bf l}_{2} + \frac{1}{16}({\bf
s}_{1}\cdot {\bf s}_{2})({\bf l}_{1}\cdot {\bf l}_{2}) - \frac{15}{64}h_{2}
- \frac{1}{32}h_{2}^{2}} \\
{\displaystyle \quad \,\, + \frac{1}{64}\big [6h_{1}({\bf l}_{1}\cdot {\bf
l}_{2}) + 4h_{1}^{2}({\bf l}_{1}\cdot {\bf l}_{2}) - 14h_{1}^{2}({\bf
l}_{1}\cdot {\bf l}_{2})^{2} + \text{H.c.}\big ],} \\
{\displaystyle h_{4} = \frac{3}{4} - \frac{7}{8}{\bf s}_{1}\cdot {\bf
s}_{2} + \frac{1}{4}{\bf l}_{1}\cdot {\bf s}_{2} + \frac{1}{4}\big [({\bf
l}_{1}\cdot {\bf s}_{1})({\bf l}_{1}\cdot {\bf s}_{2}) + \text{H.c.}\big
],} \\
{\displaystyle h_{5} = \frac{3}{4} - \frac{7}{8}{\bf s}_{1}\cdot {\bf
s}_{2} + \frac{1}{4}{\bf l}_{2}\cdot {\bf s}_{1} + \frac{1}{4}\big [({\bf
l}_{2}\cdot {\bf s}_{2})({\bf l}_{2}\cdot {\bf s}_{1}) + \text{H.c.}\big
].} \\
\end{array}
\label{e54}
\end{eqnarray}

The cell Hamiltonian $H_{1,3}$ has the same form (\ref{e53}) but with a
change of notation according to Fig.~\ref{f3}:
%Eq(s). (55)
\begin{eqnarray}
\left \{
\begin{array}{c}
{\bf l}_{1} = {\bf s}_{1}(1) + {\bf s}_{2}(1) + {\bf s}_{3}(1), \\
{\bf l}_{2} = {\bf s}_{1}(3) + {\bf s}_{3}(3) + {\bf s}_{4}(3), \\
\end{array} \right . \qquad
\left \{
\begin{array}{c}
{\bf s}_{1} = {\bf s}_{4}(1), \\
{\bf s}_{2} = {\bf s}_{2}(3). \\
\end{array} \right .
\label{e55}
\end{eqnarray}

Of special interest is the case corresponding to $\alpha = \pm \pi /2$.
Unfortunately, exact expressions for the correlation function cannot be
obtained in this case, but the Hamiltonian can be written in explicit form.
Since the site spinor $\Psi ^{\lambda \mu \nu \rho }$ does not contain a
quintet component for $\alpha = \pm \pi /2$, the wave function of two
nearest neighbor sites (\ref{e24}) and (\ref{e25}) will lack a component
with $S = 3$.  A more detailed analysis shows that 19 multiplets are
present in the wave function of two nearest neighbor sites.  In this case,
therefore, the cell Hamiltonian has the general form
%Eq(s). (56)
\begin{eqnarray}
H_{1,2} = \sum \limits _{k = 1}^{51}\lambda _{k}P_{k}^{1,2}.
\label{e56}
\end{eqnarray}

For a definite choice of $\lambda _{k}$ in Eq.~(\ref{e56}) the cell
Hamiltonian assumes the form
%Eq(s). (57)
\begin{eqnarray}
H_{1,2} &=& P_{2}({\bf l}_{1} + {\bf s}_{1}) + P_{2}({\bf l}_{2} + {\bf
s}_{2}) + P_{1/2}({\bf l}_{1})P_{1/2}({\bf l}_{2})P_{1}({\bf s}_{1} + {\bf
s}_{2}) \nonumber \\
&+& P_{3/2}({\bf l}_{1})P_{1/2}({\bf l}_{2})P_{1/2}({\bf l}_{1} + {\bf s}_{1}
+ {\bf s}_{2}) + P_{3/2}({\bf l}_{2})P_{1/2}({\bf l}_{1})P_{1/2}({\bf l}_{2}
+ {\bf s}_{2} + {\bf s}_{1}),
\label{e57}
\end{eqnarray}
where the notations (\ref{e52}) and (\ref{e55}) are used for nearest
neighbor sites along the horizontal and along the vertical, respectively.

Our results suggest that the spin correlation functions decay exponentially
with a correlation length $\sim 1$ for an arbitrary parameter $\alpha $.
We also assume that the decay of the correlation function is of the
exponential type for the 14-parameter model as well, i.e., for any choice
of site spinor $\Psi ^{\lambda \mu \nu \rho }$.  This assumption is
supported in special cases: 1)~the partition of the system into
one-dimen\-sional chains with exactly known exponentially decaying
correlation functions; 2)~the two-dimen\-sional AKLT model, for which the
exponential character of the decay of the correlation function has been
rigorously proved.\cite{14}  Further evidence of the stated assumption lies in
the numerical results obtained for various values of the parameter in the
one-parameter model.

\section[]{Generalization of the model to other types of
lattices}\label{s5}

The wave function (\ref{e7}),\,\,(\ref{e19}) can be generalized to any type
of lattice.  The general principle of wave function construction for a
system of spins~1/2 entails the following:

1) Each bond on a given lattice has associated with it two indices running
through the values 1 and 2, one at each end of the bond.

2) Each bond has associated with it a metric spinor $g_{\lambda \mu }$ with
the indices of the ends of this bond.

3) Each site of the lattice (a site being interpreted here, of course, in
the same sense as in Sec.~\ref{s3}) with {\it m} outgoing bonds has
associated with it an {\it m}th-rank spinor with the indices of the bonds
adjacent to the site.

4) The wave function is the product of all spinors at sites of the lattice
and all metric spinors.

It is obvious that each index in the formulated wave function is
encountered twice, so that the wave function is scalar and, hence, singlet.

The wave function so constructed describes a system in which each lattice
site contains as many spins $s = 1/2$ as the number of bonds emanating from
it.

To completely define the wave function, it is necessary to determine the
specific form of all site spinors.  The coefficients that determine their
form are then parameters of the model.

The Hamiltonian of such a model is the sum of the cell Hamiltonians acting
in the spin space of the subsystem formed by the spins at two mutually
coupled sites:
%Eq(s). (58)
\begin{eqnarray}
H = \sum \limits _{\langle ij\rangle }H_{ij}.
\label{e58}
\end{eqnarray}

Each cell Hamiltonian is the sum of the projectors with arbitrary positive
coefficients onto all multiplets possible in the corresponding two-site
subsystem except those present in the constructed wave function:
%Eq(s). (59)
\begin{eqnarray}
H_{i,j} = \sum \limits _{k}\lambda _{k}P_{k}^{i,j}.
\label{e59}
\end{eqnarray}
Then $H_{i,j}|\Psi _{s}\rangle = 0$ and, accordingly, $H|\Psi _{s}\rangle =
0$.

Consequently, $\Psi _{s}$ is an exact ground-state wave function.

\begin{figure}[t]
\unitlength1cm
\begin{picture}(11,6)
\centerline{\psfig{file=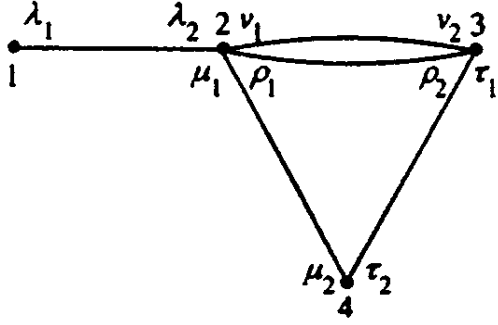,angle=0,width=8cm}}
\end{picture}
%\vspace{-2cm}
\caption[]{Example of a graph corresponding to the wave
function~(\ref{e60}).\label{f9}}
\end{figure}

We note that any two lattice sites can be joined by two, three, or more
bonds, because this does not contradict the principle of construction of
the wave function.  Moreover, the general principle of construction of the
wave function is valid not only for translationally symmetric lattices, but
for any graph in general.  As an example, let us consider the system shown
in Fig.~\ref{f9}.  The wave function of this system has the form
%Eq(s). (60)
\begin{eqnarray}
\Psi _{s} = \Psi ^{\lambda _{1}}(1)\Psi ^{\lambda _{2}\mu _{1}\nu _{1}\rho
_{1}}(2)\Psi ^{\rho _{2}\nu _{2}\tau _{1}}(3)\Psi ^{\mu _{2}\tau _{2}}(4)
g_{\lambda _{1}\lambda _{2}}g_{\mu _{1}\mu _{2}}g_{\nu _{1}\nu _{2}}g_{\rho
_{1}\rho _{2}}g_{\tau _{1}\tau _{2}}
\label{e60}
\end{eqnarray}
and describes a system containing ten spins~1/2.

If the given lattice has dangling bonds (as occurs for systems with open
boundary conditions), the resulting wave function represents a spinor of
rank equal to the number of loose ends.  The ground state of this kind of
system is therefore $2^{l}$-fold degenerate (where {\it l} is the number of
loose ends).  For an open one-dimen\-sional chain, for example, the ground
state corresponds to four functions --- one singlet and three triplet
components.  For higher-dimen\-sional lattices this degeneracy depends on
the size of the lattice and increases exponentially as its boundaries grow.

\section[]{Conclusion}\label{s6}

We have proposed a method for the construction of an exact wave function
for a class of two-dimen\-sional spin models.  In general this model
depends on 14 parameters, and its Hamiltonian is written as the sum of the
Hamiltonians of nearest neighbor spin quartets.  The exact ground-state
wave function of the total system is also the exact wave function of each
cell Hamiltonian.  Since 20 of the 70 multiplets of two nearest neighbor
quartets are present in the exact wave function, the cell Hamiltonians are
the sums of the projectors with positive coefficients onto the other 50
multiplets.  These coefficients are the excitation energies of the
corresponding multiplets.  Different values of the coefficients correspond
to different Hamiltonians.  In this case, however, the ground-state wave
function itself and the spin correlation functions in the ground state are
identical for all Hamiltonians.  This means that the ground-state wave
function, as defined by Shastry and Sutherland,\cite{18} is superstable.

We have carried out Monte Carlo calculations of the spin correlation
functions in the ground state for the special case of a model that depends
on one parameter.  For all values of the parameter the spin correlation
functions decay exponentially with distance despite the complicated
dependence of the correlation functions of nearest neighbor spins on the
model parameter.  It is justifiable to expect the spin correlations to
decay exponentially in the general 14-parameter model as well.

In closing, we are pleased to acknowledge Prof. M.~Ya. Ovchinnikova for
helpful discussions of the problems treated in the article.  This work has
received financial support from the Russian Foundation for Basic Research
(Grants No. 96-03-32186 and No. 97-03-33727) and from the Program for
Support of Leading Scientific Schools (Grant No. 96-15-97492).

\setcounter{equation}{0}
\renewcommand{\theequation}{A.\arabic{equation}}
\section*{Appendix}

In Sec.~\ref{s3} we have constructed the singlet wave function
%Eq(s). (A.1)
\begin{eqnarray}
\Psi _{s} = \prod \limits _{{\bf n}}\Psi ^{\lambda _{{\bf n}}\mu _{{\bf
n}}\nu _{{\bf n}}\rho _{{\bf n}}}({\bf n})
g_{\nu _{{\bf n}}\lambda _{{\bf n} + {\bf a}}}
g_{\rho _{{\bf n}}\mu _{{\bf n} + {\bf b}}}
\label{ea1}
\end{eqnarray}
for a system of $4M^{2}$ spins $s = 1/2$ on a square lattice.  The
following Hamiltonian was specially chosen for the resulting wave function:
%Eq(s). (A.2)
\begin{eqnarray}
H = \sum \limits _{{\bf n}}H_{{\bf n},{\bf n} + {\bf a}} +
\sum \limits _{{\bf n}}H_{{\bf n},{\bf n} + {\bf b}},
\label{ea2}
\end{eqnarray}
for which the wave function (\ref{ea1}) is the zero-energy ground state:
%Eq(s). (A.3)
\begin{eqnarray}
H|\Psi _{s}\rangle = 0.
\label{ea3}
\end{eqnarray}

We now show that the ground state of the system is nondegenerate, i.e., the
wave function satisfying Eq.~(\ref{ea3}) is unique.

Inasmuch as the Hamiltonian (\ref{ea2}) is a sum of nonnegative definite
cell Hamiltonians, any function satisfying Eq.~(\ref{ea3}) must satisfy all
the cell equations
%Eq(s). (A.4)
\begin{eqnarray}
H_{i,j}|\Psi _{s}\rangle = 0.
\label{ea4}
\end{eqnarray}
This means that Eqs.~(\ref{ea3}) and (\ref{ea4}) are equivalent.

We prove the nondegeneracy of the ground state of the Hamiltonian
(\ref{ea2}) as follows.  We first write the general form of the wave
function for the system in question.  We then determine the general form of
the wave function satisfying one of the cell equations (\ref{ea4}).  Making
note of the conditions imposed on the general form of the wave function by
each cell equation and, at the same time, simultaneously satisfying these
conditions for all the cell equations, we obtain the general form of the
wave function satisfying all the equations (\ref{ea4}) and, hence,
satisfying the total Hamiltonian~(\ref{ea2}).

Any wave function of the given system can be written in the form
%Eq(s). (A.5)
\begin{eqnarray}
\Psi = \sum \limits _{\lambda \mu \nu \rho }c(\lambda \mu \nu \rho )\cdot
\prod \limits _{j}\Phi ^{\lambda _{j}\mu _{j}\nu _{j}\rho _{j}}({\bf j}),
\label{ea5}
\end{eqnarray}
where the summation is over the $4M^{2}$ indices $\lambda _{i},\, \mu
_{i},\, \nu _{i},\, \rho _{i}$, $c(\lambda \mu \nu \rho )$ denotes
coefficients that depend on these indices, and $\Phi ^{\lambda _{j}\mu
_{j}\nu _{j}\rho _{j}}({\bf j})$ are arbitrary fourth-rank site spinors (in
general, spinors at different sites can differ).

We require that the wave function (\ref{ea5}) obey the cell equation
%Eq(s). (A.6)
\begin{eqnarray}
H_{{\bf n},{\bf n} + {\bf a}}|\Psi \rangle = 0.
\label{ea6}
\end{eqnarray}

By the construction of the singlet wave function (\ref{ea1}), which is
matched by the cell Hamiltonian $H_{{\bf n},{\bf n} + {\bf a}}$, any wave
function at sites {\bf n} and ${\bf n} + {\bf a}$ that satisfies condition
(\ref{ea6}) is a linear combination of the 64 functions contained in the
expression
%Eq(s). (A.7)
\begin{eqnarray}
\Psi ^{\lambda _{{\bf n}}\mu _{{\bf n}}\nu _{{\bf n}}\rho _{{\bf n}}}({\bf
n})
\Psi ^{\lambda _{{\bf n} + {\bf a}}\mu _{{\bf n} + {\bf a}}\nu _{{\bf n} +
{\bf a}}\rho _{{\bf n} + {\bf a}}}({\bf n} + {\bf a})
g_{\nu _{{\bf n}}\lambda _{{\bf n} + {\bf a}}},
\label{ea7}
\end{eqnarray}
because the cell Hamiltonian $H_{{\bf n},{\bf n} + {\bf a}}$ by definition
is the sum of the projectors onto all multiplets [$\Psi ^{\lambda _{{\bf
n}}\mu _{{\bf n}}\nu _{{\bf n}}\rho _{{\bf n}}}({\bf n})$ and $\Psi
^{\lambda _{{\bf n} + {\bf a}}\mu _{{\bf n} + {\bf a}}\nu _{{\bf n} + {\bf
a}}\rho _{{\bf n} + {\bf a}}}({\bf n} + {\bf a})$ are definite site spinors
occurring in the wave function (\ref{ea1})].  We note that these 64
functions can be linearly dependent (as is the case, for example, for the
two-dimen\-sional AKLT model).

Thus, the general form of the wave function satisfying Eq.~(\ref{ea6}) can
be written
%Eq(s). (A.8)
\begin{eqnarray}
\begin{array}{l}
{\displaystyle \Psi = \sum \limits _{\lambda \mu \nu \rho }c(\lambda \mu
\nu \rho |
\nu _{{\bf n}}\lambda _{{\bf n} + {\bf a}})g_{\nu _{{\bf n}}\lambda _{{\bf
n} + {\bf a}}}
\Psi ^{\lambda _{{\bf n}}\mu _{{\bf n}}\nu _{{\bf n}}\rho _{{\bf n}}}({\bf
n})} \\
\quad {\displaystyle \times \Psi ^{\lambda _{{\bf n} + {\bf a}}\mu _{{\bf
n} + {\bf a}}\nu _{{\bf n} + {\bf a}}\rho _{{\bf n} + {\bf a}}}({\bf n} +
{\bf a})\prod \limits _{{\bf j} \neq {\bf n},{\bf n} + {\bf a}}\Phi
^{\lambda _{{\bf j}}\mu _{{\bf j}}\nu _{{\bf j}}\rho _{{\bf j}}}({\bf j}),}
\\
\end{array}
\label{ea8}
\end{eqnarray}
where $c(\lambda \mu \nu \rho |\nu _{{\bf n}}\lambda _{{\bf n} + {\bf a}})$
are coefficients that depend on the indices $\lambda _{i},\, \mu _{i},\,
\nu _{i},\, \rho _{i}$ exclusive of the indices $\nu _{{\bf n}}$ and
$\lambda _{{\bf n} + {\bf a}}$, and $\Phi ^{\lambda _{j}\mu _{j}\nu
_{j}\rho _{j}}({\bf j})$ are arbitrary site spinors.

Comparing the functions (\ref{ea5}) and (\ref{ea8}), we deduce the
following conditions that must be met by the function (\ref{ea5}) to obtain
the general form of the wave function satisfying Eq.~(\ref{ea6}):

1.  The spinors at sites {\bf n} and ${\bf n} + {\bf a}$ must coincide with
the site spinors of the wave function~(\ref{ea1}):
%Eq(s). (A.9)
\begin{eqnarray}
\begin{array}{c}
{\displaystyle
\Phi ^{\lambda _{{\bf n}}\mu _{{\bf n}}\nu _{{\bf n}}\rho _{{\bf n}}}
({\bf n}) =
\Psi ^{\lambda _{{\bf n}}\mu _{{\bf n}}\nu _{{\bf n}}\rho _{{\bf n}}}
({\bf n}),} \\
{\displaystyle
\Phi ^{\lambda _{{\bf n} + {\bf a}}\mu _{{\bf n} + {\bf a}}\nu _{{\bf n} +
{\bf a}}\rho _{{\bf n} + {\bf a}}}({\bf n} + {\bf a}) =
\Psi ^{\lambda _{{\bf n} + {\bf a}}\mu _{{\bf n} + {\bf a}}\nu _{{\bf n} +
{\bf a}}\rho _{{\bf n} + {\bf a}}}({\bf n} + {\bf a}).} \\
\end{array}
\label{ea9}
\end{eqnarray}

2.  The coefficients $c(\lambda \mu \nu \rho )$ have the form
%Eq(s). (A.10)
\begin{eqnarray}
c(\lambda \mu \nu \rho
) = c(\lambda \mu \nu \rho |\nu _{{\bf n}}\lambda _{{\bf n} + {\bf
a}})g_{\nu _{{\bf n}}\lambda _{{\bf n} + {\bf a}}}.
\label{ea10}
\end{eqnarray}

From the equation
%Eq(s). (A.11)
\begin{eqnarray}
H_{{\bf n},{\bf n} + {\bf b}}|\Psi \rangle = 0
\label{ea11}
\end{eqnarray}
we deduce analogous conditions on the general form of the wave
function~(\ref{ea5}):
%Eq(s). (A.12)
\begin{eqnarray}
\Phi ^{\lambda _{{\bf n}}\mu _{{\bf n}}\nu _{{\bf n}}\rho _{{\bf n}}}({\bf
n}) &=& \Psi ^{\lambda _{{\bf n}}\mu _{{\bf n}}\nu _{{\bf n}}\rho _{{\bf
n}}}({\bf n}), \nonumber \\
\Phi ^{\lambda _{{\bf n} + {\bf b}}\mu _{{\bf n} + {\bf b}}\nu _{{\bf n} +
{\bf b}}\rho _{{\bf n} + {\bf b}}}({\bf n} + {\bf b}) &=&
\Psi ^{\lambda _{{\bf n} + {\bf b}}\mu _{{\bf n} + {\bf b}}\nu _{{\bf n} +
{\bf b}}\rho _{{\bf n} + {\bf b}}}({\bf n} + {\bf b}), \label{ea12} \\
c(\lambda \mu \nu \rho ) &=&
c(\lambda \mu \nu \rho |\rho _{{\bf n}}\mu _{{\bf n} + {\bf
b}})g_{\rho _{{\bf n}}\mu _{{\bf n} + {\bf b}}}. \nonumber
\end{eqnarray}

The simultaneous satisfaction of all the cell equations (\ref{ea4})
requires consolidation of the conditions imposed by these equations on the
general form of the wave function (\ref{ea5}).  Combining these conditions
in succession, in each step we obtain the general form of a wave function
satisfying the equations corresponding to these conditions.  Upon
satisfying all the conditions, we obtain the general form of the wave
function satisfying all the cell equations (\ref{ea4}) and, hence,
satisfying Eq.~(\ref{ea3}):
%Eq(s). (A.13)
\begin{eqnarray}
\Psi _{s} = \sum \limits _{\lambda \mu \nu \rho }c(\lambda \mu \nu \rho
|\lambda \mu \nu \rho )\prod \limits _{{\bf j}}\Psi ^{\lambda _{{\bf j}}\mu
_{{\bf j}}\nu _{{\bf j}}\rho _{{\bf j}}}({\bf j})
g_{\nu _{{\bf j}}\lambda _{{\bf j} + {\bf a}}}
g_{\rho _{{\bf j}}\mu _{{\bf j} + {\bf b}}},
\label{ea13}
\end{eqnarray}
where $c(\lambda \mu \nu \rho |\lambda \mu \nu \rho ) = c$ is a constant.

Comparing the wave functions (\ref{ea1}) and (\ref{ea13}), we readily
perceive that, to within an arbitrary factor, the general form of the wave
function satisfying Eq.~(\ref{ea3}) coincides with the wave function $\Psi
_{s}$.  Consequently, $\Psi _{s}$ is the nondegenerate ground-state wave
function of the Hamiltonian~(\ref{ea2}).

%\begin{flushleft}
%Translated by James S. Wood
%\end{flushleft}

\end{document}